%% file: main.tex
\documentclass[runningheads]{llncs}
\usepackage[T1]{fontenc}
\usepackage{graphicx}
\usepackage{graphicx}
\usepackage{algorithm}
\usepackage{algorithmic}
\usepackage{subcaption}
\usepackage{comment}
\usepackage{wrapfig}
\usepackage{amsmath}
\usepackage{amssymb}
\usepackage[hidelinks]{hyperref}  

\newcommand\blfootnote[1]{%
  \begingroup
  \renewcommand\thefootnote{}\footnote{#1}%
  \addtocounter{footnote}{-1}%
  \endgroup
}

\begin{document}
%
\title{Maximizing the efficiency of human feedback in AI alignment: a comparative analysis}
\titlerunning{Maximizing the efficiency of human feedback in AI alignment}
%
\author{Andreas Chouliaras \and Dimitris Chatzopoulos}
\authorrunning{A.Chouliaras and D.Chatzopoulos.}
%
\institute{University College Dublin, Ireland \\
    \email{andreas.chouliaras@ucdconnect.ie}, \\
    \email{dimitris.chatzopoulos@ucd.ie}
}
\maketitle
\begin{abstract}
Reinforcement Learning from Human Feedback (RLHF) relies on preference modeling to align machine learning systems with human values, yet the popular approach of random pair sampling with Bradley–Terry modeling is statistically limited and inefficient under constrained annotation budgets. In this work, we explore alternative sampling and evaluation strategies for preference inference in RLHF, drawing inspiration from areas such as game theory, statistics, and social choice theory. Our best-performing method, Swiss InfoGain, employs a Swiss tournament system with a proxy mutual-information-gain pairing rule, which significantly outperforms all other methods in constrained annotation budgets while also being more sample-efficient. Even in high-resource settings, we can identify superior alternatives to the Bradley–Terry baseline. Our experiments demonstrate that adaptive, resource-aware strategies reduce redundancy, enhance robustness, and yield statistically significant improvements in preference learning, highlighting the importance of balancing alignment quality with human workload in RLHF pipelines.


\end{abstract}

\section{Introduction}
Reinforcement\blfootnote{\textbf{Extended version with appendix:} \url{https://arxiv.org/abs/2511.12796}} Learning from Human Feedback (RLHF) has emerged as a central paradigm for aligning machine learning systems with human preferences~\cite{christiano2017deep,ibarz2018reward,ziegler2019fine,ouyang2022training}. RLHF leverages human preference judgments to guide policy optimization, typically by fitting a reward model from pairwise comparisons of model outputs. The efficiency of this pipeline critically depends on how sample pairs are generated and how the resulting comparisons are aggregated. In the standard approach, pairs are generated at random and modeled using a Bradley–Terry framework, which assumes latent utility scores and fits them via maximum likelihood estimation~\cite{christiano2017deep}. While statistically principled, this approach does not account for annotation cost, and its sample efficiency is limited under constrained human evaluation budgets~\cite{casper2023open,Wang2024ACS}. Notably, training the Bradley-Terry model with random pairs of samples hasn't been contested since its first adoption by Christiano et al.~\cite{christiano2017deep}. 
This raises the following critical question: 

\textit{How to better make use of human effort to create accurate and efficient reward models across different annotation budgets?}

In the low-resource regime, random sampling often produces redundant or uninformative comparisons, thereby reducing the effective information gain per annotation. Alternative strategies, such as adaptive sampling, active ranking algorithms, or methods inspired by statistics, social choice theory, and game theory, provide a promising alternative in order to mitigate this inefficiency by selecting comparisons that are expected to maximize model improvement. These methods leverage structural properties of the preference model and uncertainty estimates to better allocate scarce annotation resources. Despite promising theoretical guarantees and empirical results in other areas of research, a systematic evaluation of their applicability to RLHF and their potential relative performance across feedback regimes is still lacking.

This paper addresses this gap by analyzing and comparing different sampling and evaluation strategies for preference modeling in RLHF under varying availability for human feedback. We also identify conditions under which traditional random sampling with Bradley–Terry modeling becomes sub-optimal. Our results show that when human resources are limited, reliance on random sampling leads to sub-optimal preference inference, while adaptive strategies yield significant gains in efficiency and robustness\footnote{Code available at: \url{https://github.com/achouliaras/aics2025_rlhf_sampling}}. These findings highlight the importance of incorporating resource-aware design principles into RLHF pipelines that better balance the competing demands of model alignment and human workload.

\section{Background}
In RLHF, one of the objectives is to generate examples of the RL systems' use, pair them, and show them to human critics to evaluate. In the case of binary preferences - which is the most popular form of human feedback and is backed by studies in cognitive sciences~\cite{hilton1986knowledge} - the process involves pairing these RL examples (referred to as items for brevity). Comparing between two items isn't always deterministic and usually leads to different evaluations from different human critics~\cite{casper2023open,dai2024mapping}. One explanation for this is that in most applications (e.g, LLM responses) the preference between two outputs hides a quite extensive evaluation of qualitative variables. But this doesn't happen for every pair of items. For some pairs, the users can point at the better one with great confidence, while for more similar pairs of items, the distinction is less apparent. 

\begin{figure}[h]
    \centering
    \includegraphics[width=\columnwidth]{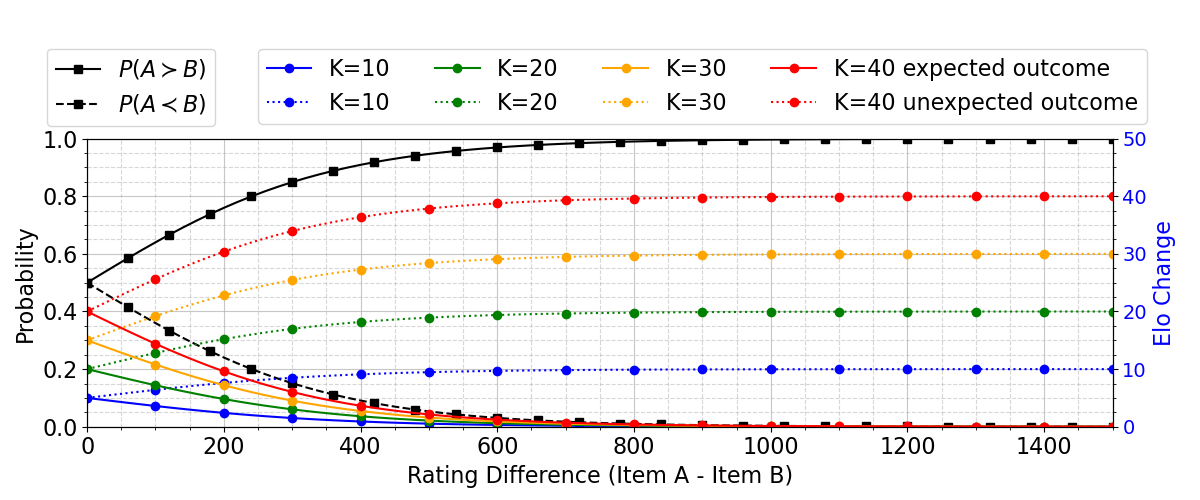}
    \caption{Elo change based on initial rating difference for various K-Factor values.}
    \label{fig:k_factor}
\end{figure}

This implies an inherent latent value score $v(x)$ that we can't measure but only estimate. These estimations are noisy due to: (i) the subjective criteria human critics often rely on, (ii) the difference in importance of such criteria between different critics, and (iii) poor perception of the differences between two items.
Every method that tries to estimate the latent true value $v(x)$ can add even more to this stochasticity. Considering $N$ items ($\mathcal{X}=[x_1,x_2,...,x_N$]) that need to be evaluated, such that $|\mathcal{X}|= N$, we define as $v(x)$ the latent true value of each item such that considering two items $x_i, x_j$, then $x_i \succ x_j \Leftrightarrow v(x_i)>v(x_j)$, where  $x_i \succ x_j$ means $x_1$ is more preferable to $x_2$. To model the probability $P(x_i \succ x_j)$ the Bradley-Terry model~\cite{Bradley1952RankAO} is used that defines $P$ as:
\begin{equation}
    P(x_i \succ x_j) = \frac{e^{v(x_i)}}{e^{v(x_i)}+e^{v(x_j)}}
    \label{eq:BT}
\end{equation}
Getting inspiration from the Elo rating system~\cite{elo1978rating}  that uses a scale factor of 400 we can transform equation~\ref{eq:BT} to:
\begin{equation}
    P(x_i \succ x_j) = \frac{1}{1 +e^{(v(x_j)-v(x_i))/400}}
    \label{eq:Elo}
\end{equation}
When the distributions of the item values ($V\sim \mathcal{N}(\mu, \sigma)$) are scaled to $\mu=1000$ and $\sigma=200$, then the Elo rating system states that: a 100 point difference between two items results in 64\% chances of the best one to win, 200 point difference results in 72\% and for 400 point difference results to around 90\%. Also, the Elo rating system provides a score update capped by a $K$-factor that is determined by the number of times an item was compared with other items:
\begin{equation}
   \hat{v}(x_i)  = \hat{v}(x_i) + K(S(x_i)-E(x_i)) 
    \label{ELO update}
\end{equation}
where $\hat{v}(x_i)$ is the estimated value of item $x_i$, $E(x_i)$ is the expected outcome of comparing $x_i$ with other items, based on equation~\eqref{eq:Elo}, and $S(x_i)$ is the actual outcome of these comparisons (defined as 1 for wins, 0.5 for ties and 0 for losses). Figure \ref{fig:k_factor} shows how different K values limit the Elo changes as the rating difference becomes progressively bigger.

\section{Methodology}
In this section, we present the methods we use in our analysis inspired by areas such as statistics~\cite{Bradley1952RankAO}, game theory~\cite{hua2017swiss}, and social choice theory~\cite{mclean1995classics,dai2024mapping}. We initially explore the performance of the Bradley-Terry estimator model, the most popular method used for getting human preferences in RLHF. Motivated by its inefficiencies in the random-pairs sampling approach, we observed that its performance takes a significant drop for constrained annotation budgets. This set us out to explore other sampling alternatives that seek to create pairs that yield optimal results with albeit fewer data samples. From this endeavor, we develop two broad families of methods. The first one relies on the \emph{Borda count method}, which ranks items based on the number of times they win over other items, using two different sampling strategies. The second one builds upon the Swiss tournament method as a way to further optimize pair sampling in an even more informed way. Standing out among these methods is our algorithm named \emph{Mutual Information gain pairs in Swiss tournament (Swiss InfoGain)} that outperforms all other methods, including the current state of the art estimator (Bradley-Terry) in both efficiency and prediction performance.

\subsection{The Bradley-Terry estimator.} 
This method uses a parametric version of Eq. \eqref{eq:BT} to fit an ML estimator $f$ such that the estimated value of an item is $\hat{v}_i = f(x_i)$. The estimator can be more advanced depending on the number of parameters needed to describe $x_i$. In our experiments, we use Algorithm \ref{algo:bt} which rescales the ratings such that $\hat{v} \sim \mathcal{N}(1000,200)$. The Bradley-Terry estimator relies in pairs sampled randomly (N/2 comparisons when redundancy $c = 1$).

\subsection{The Borda count method.} 
\begin{figure}[ht]
    \centering
    \includegraphics[width=0.65\columnwidth]{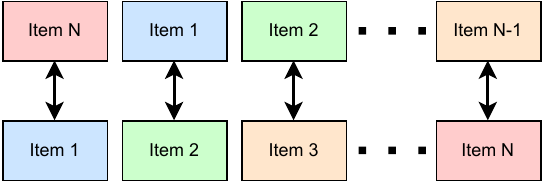}
    \caption{The Copeland pairing algorithm.}
    \label{fig:copeland}
\end{figure}

The Borda count method comes from the social choice theory and offers a simple yet effective positional voting rule where the score of each item is the number of wins it has at the end~\cite{mclean1995classics}. We apply this method in our tests with two different sampling strategies: either randomly ($N/2$ comparisons per round) or in a round-robin style manner known as Copeland's method (Figure~\ref{fig:copeland}) in social choice theory~\cite{copeland1951reasonable,saari1996copeland} when the resources are unrestricted (requires $N(N-2)/2$ comparisons per round). We refer the readers to Algorithm~\ref{algo:borda} for our implementation of the random sampling approach, with the Copeland variation being very similar. We also tried to implement the Quicksort approach based on~\cite{maystre2017quicksort} and assigning a Borda count score after each sorting round. The results for this technique were much poorer than the methods included in our experiments and, therefore, are omitted from the figures in our experiments later on, but it will be included, however, in our code-base release.

\subsection{The Elo rating system}
This method applies the Elo rating system Eqs.~\eqref{eq:Elo} \& \eqref{ELO update}. We chose a simplified form instead of variations that are designed for specific applications (e.g, chess). The Elo can be updated either after each comparison or at the end of the round, with no differences in the results. The pairs can be either sampled randomly or using the Copeland method (Figure~\ref{fig:copeland}). Refer to Algorithm~\ref{algo:elo_rnd} for the random sampling approach, with the Copeland variation being very similar.

\subsection{The Swiss tournament method}
\begin{figure}[ht]
    \centering
    \includegraphics[width=0.75\columnwidth]{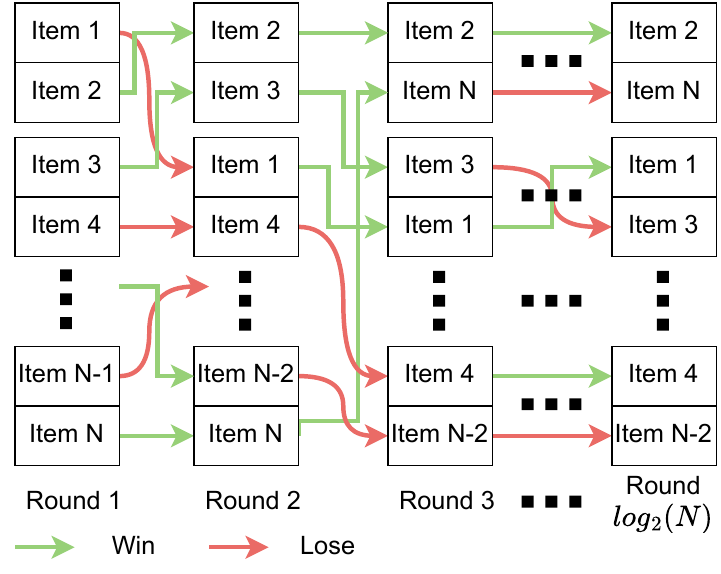}
    \caption{The Swiss tournament system.}
    \label{fig:swiss}
\end{figure}

The Swiss tournament system involves pairing items based on their score~\cite{hua2017swiss} (Figure~\ref{fig:swiss}). In each round, every item is paired with the item with the closest running score. At the end of each round, all items update their Elo ratings and are paired again with other items in the next round~\cite{hua2017swiss} (Algorithm~\ref{algo:swiss_elo}).

\subsection{Random pairs then Swiss Tournament.} The Swiss tournament system -- created for chess tournaments and widely applied and studied by game theory -- involves pairing items based on their score~\cite{hua2017swiss}. In each round, every item is paired with the item with the closest running score. At the end of each round, all items update their Elo ratings and are paired again with other items in the next round~\cite {hua2017swiss}. While this method guarantees the best items come to the top, and the worst go to the bottom, it leaves the middle point a little more clouded. Furthermore, when every item is originally unranked, applying it in its pure form can lead to discretised distributions of scores. We implement two variations of this method, one being the pure Swiss tournament method, with the other having random pairings in the beginning and then applying the Swiss method. Algorithm~\ref{algo:rnd_swiss} describes the second approach, while setting the random rounds to zero, it regresses back to the pure Swiss tournament method. The reason for exploring random pairing rounds in the beginning is that the Elo system is path dependent, meaning that the order in which an item compares with others affects the final resulting rating. With all items having a more noisy initial rating, applying the Swiss method can further separate the items even faster. Items belonging in the middle of the distribution will have greater initial variance, thus leading to potentially better separation. 
Concerning the allocation of the annotation budget between random pairing rounds and Swiss pairing rounds, we examined all combinations in our simulations with the 50-50 rule, yielding, on average, better results. This finding is convenient for its applicability in RLHF, as the randomly sampled pairs can be pre-generated and therefore substantially reduce the computational cost required during the annotation session.

\subsection{Mutual information gain pairs in Swiss tournament.}
Trying to push the potential of the Swiss tournament method even further, we test the hypothesis that pairing items from their current Elo rating might not be efficient enough in the early rounds, and random pairs might be too wasteful. This comes from the intuition that the running Elo rating improves its accuracy with successive pairings, but its path dependence remains problematic. We wanted to test the Swiss tournament in a form without relying directly on the Elo system or the initial random pairs, as in the previous method. We chose to use information gain as the alternative pairing rule.  With the same initial rating, in the first round, all pairs are purely random. Based on the result, all items are then split into different bands based on their score. Instead of pairing them based on Elo rating closeness, we use information gain to also allow cross-band pairs, making for a more dynamic pairing rule. By pairing them with items that yield the greatest information in each round and discarding non-informative pairs, we gradually separate the item groups and spread them until there are no more informative pairs. If $P(x_i \succ x_j)\approx 0 \text{ or } 1$, the pairing isn't informative, whereas if $P(x_i \succ x_j)\approx 0.5$ has the greatest uncertainty in the outcome and therefore makes a more informative pair. Information gain expressed in terms of entropy is defined as $IG(x_i,x_j)=H(x_i)-H(x_i|x_j)$. The probability $p = P(Y=1|x_i,x_j) = P(x_i \succ x_j)$ can be expressed as a Bernoulli trial of a random variable $Y\in \{0,1\}$. The entropy of such variables is defined as 
\begin{equation}
    H(Y)=-plog(p)-(1-p)log(1-p) \label{eq:entropy}
\end{equation}    
We approximate $IG(x_i,x_j)\propto H(Y)$ Based on S. Shams~\cite{shams2011maximum} we can further simplify Eq.~\eqref{eq:entropy} by replacing the entropy with a second order Taylor-approximation as $IG(x_i,x_j) = p(1-p)$. As $p=P(x_i \succ x_j)$ and $1-p=P(x_i \prec x_j)$ the mutual information gain approximation becomes:
\begin{equation}
    IG(x_i,x_j) = P(x_i \succ x_j) \cdot P(x_i \prec x_j)
    \label{eq:infogain}
\end{equation}
Where $P(x_i \succ x_j)$ and $P(x_i \prec x_j)$ comes from Eq.~\eqref{eq:BT}.
The IG approximation is maximized at $p=0.5$ and gets to zero at the extremes. Refer to Algorithm~\ref{algo:swiss_infogain} for the implementation of this method.

\section{Experimental Evaluation}
In an ideal setting, a method estimating the latent values $\hat{v}$ should perfectly align with the true values $v$. In such a case, the correlation between the estimated and true values would approach one ($r(\hat{v}, v) \to 1$). Considering all methods introduced so far, in this work, we address the following research questions:
\begin{enumerate}
    \item Which method maximizes $r(\hat{v},v)$ given a limited annotation budget $M$?
    \item How do the methods perform as the annotation budget $M$ increases?
\end{enumerate}
To model this problem, we considered a set of $N$ items, each assigned a latent true value sampled from a normal distribution $\mathcal{N}(1000, 200)$. To determine the expected outcome between two items $x_a,x_b$, we first model the probability of equal preference in the following way: 

\begin{equation}
    P(x_a\sim x_b) = \frac{1}{3} e^{-|v(x_a)-v(x_b)|/\sigma}
    \label{eq:tie_prob}
\end{equation}

\begin{figure}[ht]
    \centering
    \includegraphics[width=0.75\columnwidth]{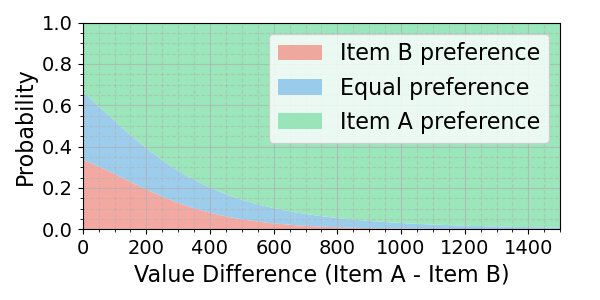}
    \caption{Probabilities of comparison outcome based on value difference}
    \label{fig:outcome_probs}
\end{figure}

When two items have very similar latent values, each possible outcome of the comparison — $x_a \succ x_b$, $x_a \sim x_b$, or $x_a \prec x_b$ — is equally likely, with probability $1/3$. Equation~\eqref{eq:tie_prob} models how the probability of a tie decreases as the difference between item values increases, as illustrated in Figure~\ref{fig:outcome_probs}. The remaining probability mass is allocated according to the Elo-style formulation in Equation~\eqref{eq:Elo}. Notably, when the difference in values is large, the probability of a tie can exceed that of the weaker item being chosen. This behavior aligns with intuition from game-theoretic settings, where parings of greatly outmatched items produce ties more frequently than surprising results.

For our experiments, we evaluated the following techniques: i) the Bradley–Terry model estimator (Bradley–Terry), ii) the Borda count method with random sampling (Borda–RNG), iii) the Borda count method combined with Copeland’s sampling strategy (Borda–Copeland), iv) Elo rating with purely random sampling (Elo–RNG), v) Elo rating combined with Copeland’s sampling strategy (Elo–Copeland), vi) a pure Swiss tournament using Elo ratings (Swiss Tournament), vii) random sampling followed by a Swiss tournament using the Elo rating system (Elo–RNG+Swiss), and viii) our Swiss tournament system variation with an information–gain–based pairing rule (Swiss–InfoGain). In addition, we explored a few other approaches (e.g the Quicksort–based ranking methods of Maystre and Grossglauser~\cite{maystre2017quicksort}) that are omitted from our reports due to poor performance or similarity to the methods covered previously in this paper.

\subsection{Comparing methods given a fixed annotation budget $M$}
\begin{figure}[ht]
    \centering
    \includegraphics[width=\columnwidth]{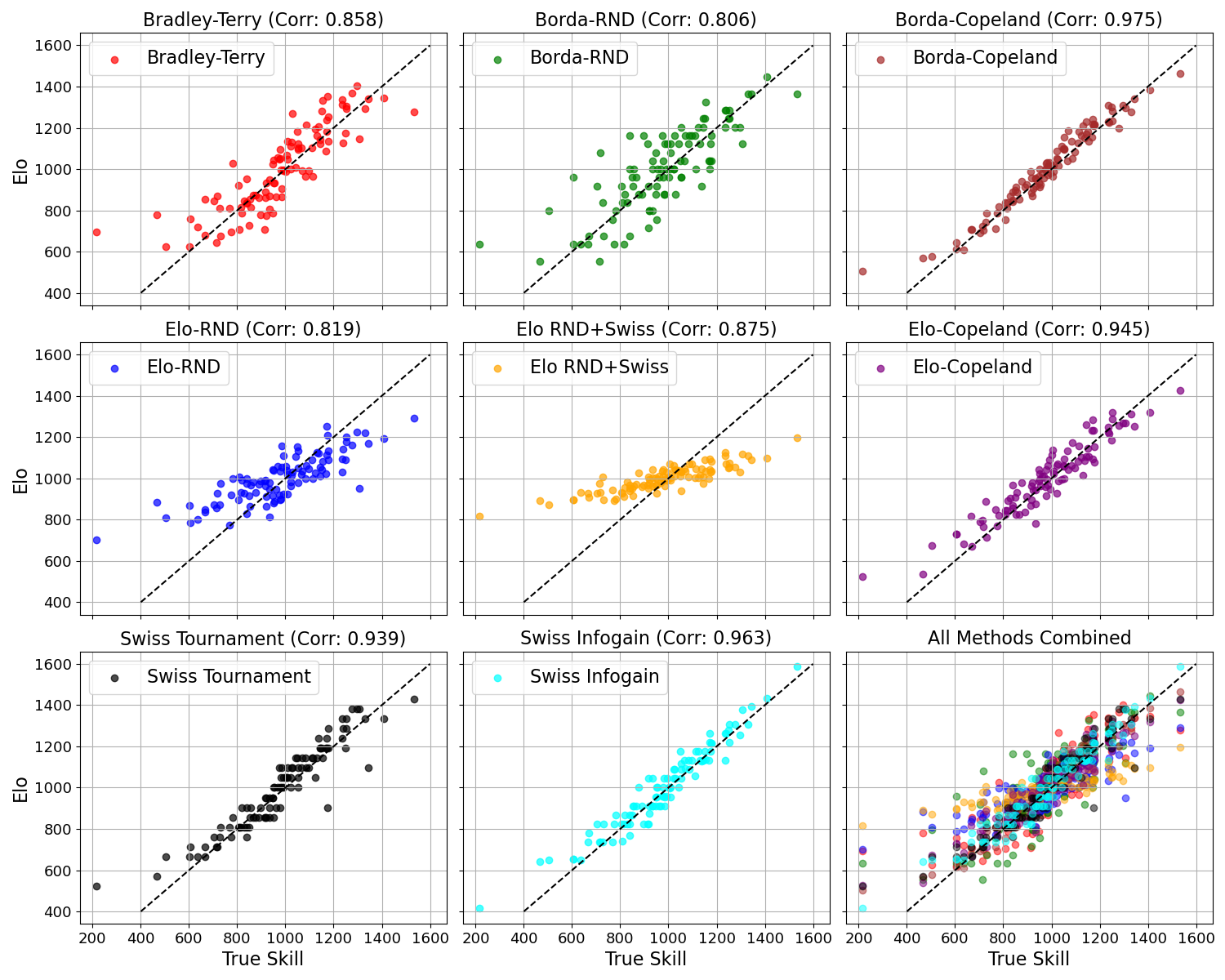}
    \caption{Correlation of Estimated value vs True value.}
    \label{fig:correlations} 
\end{figure}

To answer our first research question, we consider $N=100$ items each with a randomly assigned latent true value. For the methods that operate using the Elo system, we use an initial Elo rating of 1000, while for the non-Elo methods,  we simply scale back the results to $\mu=1000$, $\sigma=200$) distribution. We calibrated each method to use the same (or very similar) number of comparisons. The exception to that is the Borda-Copeland and Elo-Copeland methods due to the need to create all symmetrical combinations of items (4950 pairs). The other methods operate in around 550 pairs. This number was chosen because the Swiss tournament method operates with 450-550 comparisons, as it stops when no two items share the same score. The results are presented in Figure~\ref{fig:correlations}.

First, we observe that the Copeland methods display the best performance at around 0.96, but they require 9 times more comparisons compared to the other methods. They serve as the ideal scenario to compare against. In fact, when running the other methods in the same budget, they also display similar performance. The next noteworthy detail is the relatively poor performance of random sampling methods. In fact, Borda-RND, Elo RND, and Elo RND+Swiss display subpar performance compared to the Bradley-Terry estimator. Notably, only the full Swiss tournament methods were able to systematically beat the Bradley-Terry model in low annotation budgets. Especially for the case of Swiss InfoGain, it outperforms even the Copeland methods and uses 9 times less data. This performance gain, though, comes with a trade-off. The Bradley-Terry model, in the case of randomly sampled pairs, generates all pairs before the annotation task. The Swiss Tournament methods, however, need multiple rounds in order to analyze the results and suggest new pairs.

\begin{figure}[t]
    \centering
    \includegraphics[width=\columnwidth]
    {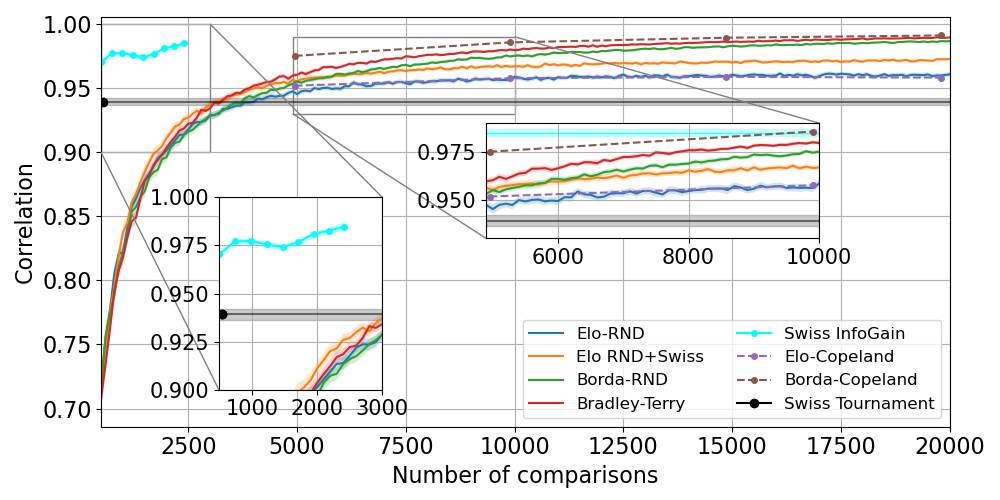}
    \caption{Comparative performance based on total number of comparisons.}
    \label{fig:perf_combined}
\end{figure}

\subsection{Comparing the method performance as $M$ increases}
At this point, the question is intuitive: What happens for other budgets? Established practices expect the Bradley-Terry model to be the best for excessive annotation budgets, but as we see, some of the methods can be better for lower budgets. The question we try to answer here is what's the tipping point, considering $N=100$ items again. In this experiment, we compare the following methods: Bradley-Terry, Borda-RND, Elo-RND, and Elo RND+Swiss by conducting the simulations from budgets of 500 all the way to 20.000 pairs. For statistical significance, we use 100 random seeds and report the mean values in bands with 95\% confidence intervals. Note, the Copeland methods can't be scaled to operate on the full range of 500 to 20.000 comparisons, so we only report the results in multiples of 4950 comparisons. As for the Swiss tournament method, it operated on average 500 pairs across the 100 seeds, and for the Swiss InfoGain method, it needs, on average, fewer than 2500 pairs. The results are in Figure~\ref{fig:perf_combined}. There are two very interesting findings here. First, the Swiss InfoGain method outperformed everything else across most of the comparison budget range (considering $N=100$). This makes it very efficient for most annotation budgets. In fact, it needs more than 15,000 pairs for the Borda-Copeland method to clearly outperform it (measuring to \~83\% greater data efficiency). For higher budgets, the Borda-Copeland method remains the clear winner, requiring a staggering 20,000 pairs for the Bradley-Terry model to match its performance. Compared to Swiss InfoGain, the Bradley-Terry model requires more than 17,500 pairs to outperform it, measuring to \~86\% greater data efficiency.
These results point towards a very interesting design regime. For low to medium annotation budgets, the Swiss InfoGain algorithm is the most efficient and potent method to create item pairs, but for unrestricted annotation budgets, the Borda-Copeland performs best.

\section{Discussion}
Our results show that, under scenarios with virtually unlimited annotation budgets, the Borda–Copeland method consistently produces stronger aggregate rankings and leads to more stable reward inference than random pair sampling. In contrast, when annotation budgets are constrained, our proposed Swiss InfoGain method generates denser, more diverse, and ultimately higher-quality annotations, thereby maximizing the information obtained per human preference. Taken together, these findings suggest that while Bradley–Terry modeling remains a necessary foundation for reward estimation, integrating alternative sampling and aggregation strategies upstream offers substantial gains in both efficiency and robustness, pointing toward more holistic, resource-aware RLHF pipelines.

\section{Limitations and comparison to related work} 
The performance and efficiency improvements demonstrated in our work primarily target the earlier stage of the pipeline, that is, how to assemble a more informative and less redundant dataset of annotated item pairs that can then be used to train reward models more effectively. The methods presented don't aim at replacing the Bradley–Terry probability model or other more recent approaches that rely on Active Learning. This limitation arises because, in RLHF, the reward model is not only trained to predict user preferences at the level of annotated item pairs but also to amortize the learned reward signal across every individual state in a trajectory, or every token in the case of large language models. This fine-grained decomposition is essential for downstream policy optimization, and the Bradley–Terry formulation remains a principled and widely applicable way to map comparisons into latent utility scores.

\section{Conclusion}
In summary, we find that random pair sampling with Bradley–Terry modeling, though widely adopted as the de facto standard in RLHF, is inefficient when the availability of human feedback is limited, as it often results in redundant or uninformative comparisons. Based on the results of our experiments, we demonstrate that our proposed method, Swiss InfoGain, which employs a Swiss tournament system combined with a mutual information gain style pairing rule, consistently achieves better performance while requiring substantially fewer annotations, thereby offering significant improvements in sample efficiency. Furthermore, we argue that even when annotation budgets are effectively unrestricted, alternative aggregation schemes such as a round-robin Borda count approach provide superior outcomes compared to the traditional Bradley–Terry model trained on randomly generated pairs, highlighting the broader applicability of resource-aware design principles. The statistical significance of these findings provides strong evidence that current RLHF pipelines must be reconsidered and restructured to incorporate more principled sampling and evaluation strategies. By doing so, it becomes possible not only to ensure stronger alignment between models and human values but also to achieve this alignment in a manner that reduces the burden on human annotators, ultimately enabling more scalable and sustainable deployment of preference-based learning systems.

\begin{credits}
\subsubsection{\ackname}
This work was supported in part by the EU Horizon projects with numbers 101092912 (MLSysOps) and 101160671 (DIGITISE).
\end{credits}

\bibliographystyle{splncs04}
\bibliography{rl, xirl}

\appendix
\include{appendix}

\end{document}

%% file: appendix.tex
\section{Algorithms}

\begin{algorithm}[ht]
\caption{Bradley--Terry Estimator (with Gradient Descent)}
\begin{algorithmic}[1]
\STATE \textbf{Input:} Number of items $N$, true values $v$, redundancy $c$
\STATE \textbf{Output:} Estimated  values $\hat{v}$, total number of comparisons $M$

\STATE Let $\mathcal{X} \gets$ set of items with $|\mathcal{X}|=N$
\STATE Initialize empty match list $m \gets [\,]$
\STATE $M \gets 0$ \hfill \COMMENT{total comparisons counter}
\FOR{$i = 1$ to $(c \cdot N / 2)$}
    \STATE $x_1, x_2 \gets sample(\mathcal{X}) \;\;|\;\; x_1\neq x_2$ 
    \STATE $(s_1, s_2) \in \{(1,0), (0.5,0.5), (0,1)\}$  w.p. $P(x_1 \succ x_2)$ Eq. \eqref{eq:BT}
    \STATE $m \gets m \cup [(x_1, x_2, s_1, s_2)]$
    \STATE $M \gets M + 1$
\ENDFOR
\STATE Initialize skill estimates $f \gets \{x : 0.0 \;\;|\;\; x \in \mathcal{X}\}$
\FOR{$t = 1$ to $20$}
    \FORALL{$(x_i, x_j, s_i, s_j) \in m$}
        \STATE $\hat{P}(x_i \succ x_j)) \gets e^{f(x_i)}/(e^{f(x_i)} + e^{f(x_j)})$ \hfill \COMMENT{Eq. \eqref{eq:BT}}
        \STATE Update gradients: 
        $\nabla_{x_i} \mathrel{+}= s_x - P(x_i \succ x_j), \quad
        \nabla_{x_j} \mathrel{+}= s_j - P(x_j \succ x_i)$
    \ENDFOR
    \STATE $f'(x) \gets f(x) + 0.01 \cdot \nabla_x$
\ENDFOR
\STATE $\hat{v} \gets (f(x) - \mu(f(x)))/\sigma(f(x)) \cdot 200 + 1000 \quad \forall x \in \mathcal{X}$
\RETURN $\hat{v}, M$
\end{algorithmic}
\label{algo:bt}
\end{algorithm}

\begin{algorithm}[h]
\caption{Borda count (with Random Matching)}
\begin{algorithmic}[1]
\STATE \textbf{Input:} Number of items $N$, true values $v$, number of rounds $r$
\STATE \textbf{Output:} Estimated values $\hat{v}$, total number of comparisons $M$

\STATE Let $\mathcal{X} \gets$ set of items with $|\mathcal{X}|=N$
\STATE Initialize wins $w(x) \gets 0$ for all $x \in \mathcal{X}$
\STATE $M \gets 0$ \hfill \COMMENT{total comparisons counter}

\FOR{$k = 1$ to $r$}
    \STATE shuffle $\mathcal{X}$ 
    \FOR{$i = 1$ to $N-1$ with step $2$}
        \STATE $x_i, x_j \gets$ $\mathcal{X}(i),\mathcal{X}(i+1)$
        \STATE $(s_i, s_j) \in \{(1,0), (0.5,0.5), (0,1)\}$  w.p. $P(x_i \succ x_j)$ {Eq.~\eqref{eq:BT}}
        \STATE $w(x_i) \gets w(x_i) + s_i$ , $w(x_j) \gets w(x_j) + s_j$
        \STATE $M \gets M + 1$
    \ENDFOR
\ENDFOR
\STATE $\hat{v} \gets (f(x) - \mu(f(x)))/\sigma(f(x)) \cdot 200 + 1000 \quad \forall x \in \mathcal{X}$
\RETURN $\hat{v}, M$
\end{algorithmic}
\label{algo:borda}
\end{algorithm}

\begin{algorithm}[ht]
\caption{Elo rating (with Random Matching)}
\begin{algorithmic}[1]
\STATE \textbf{Input:} Number of items $N$, true values $v$, initial ratings $\mathcal{R}_0$, number of rounds $r$, $K$-factor $K$
\STATE \textbf{Output:} Estimated values $y$, total number of comparisons $M$

\STATE Initialize ratings $\hat{u} \gets \mathcal{R}_0$
\STATE Let $\mathcal{X} \gets$ set of items with $|\mathcal{X}|=N$
\STATE $M \gets 0$ \hfill \COMMENT{total comparisons counter}

\FOR{$k = 1$ to $r$}
    \STATE shuffle $\mathcal{X}$
    \FOR{$i = 1$ to $N-1$ step $2$}
        \STATE $x_i, x_j \gets$ $\mathcal{X}(i),\mathcal{X}(i+1)$
        \STATE $(S_i, S_j) \in \{(1,0), (0.5,0.5), (0,1)\}$  w.p. $P(x_i \succ x_j)$ Eq.~\eqref{eq:Elo}
        \STATE $(E_i, E_j) = P(x_i \succ x_j), P(x_j \succ x_i)$ Eq.~\eqref{eq:Elo}
        \STATE $M \gets M + 1$
    \ENDFOR
    \STATE batch Elo update $\hat{u}(x) \gets \hat{u}(x) + K \cdot (S(x) - E(x)) \;\; \forall\;\; x \in \mathcal{X}$
\ENDFOR
\RETURN $\hat{u}, M$
\end{algorithmic}
\label{algo:elo_rnd}
\end{algorithm}

\begin{algorithm}[ht]
\caption{Swiss Tournament}
\begin{algorithmic}[1]
\STATE \textbf{Input:} Number of items $N$, true values $v$, initial ratings $\mathcal{R}_0$, max rounds $r_{\max}$, base $K$-factor $K_0$
\STATE \textbf{Output:} Final ratings $\mathcal{R}$, total number of comparisons $M$

\STATE Initialize ratings $\hat{u} \gets \mathcal{R}_0$
\STATE Let $\mathcal{X} \gets$ set of items with $|\mathcal{X}|=N$
\STATE $M \gets 0$, $t \gets 0$ \hfill \COMMENT{total comparisons and round counters}

\WHILE{$t < r_{\max}$}
    \STATE group items by value $s \gets \{y(x) : s(y(x))\cup [x] \in \mathcal{X}\}$ 
    \STATE Initialize empty match list $m \gets [\,]$
    \FORALL{$\hat{y}$ in $s$}
        \STATE shuffle $\hat{y}$
        \FOR{$i = 1$ to $|\hat{y}|-1$ step $2$}
            \STATE $x_i, x_j \gets$ $\hat{y}(i),\hat{y}(i+1)$
            \STATE $(S_i, S_j) \in \{(1,0), (0.5,0.5), (0,1)\}$ w.p. $P(x_i \succ x_j)$ Eq. \eqref{eq:BT}
            \STATE $(E_i, E_j) = P(x_i \succ x_j), P(x_j \succ x_i)$ Eq.~\eqref{eq:Elo}
            \STATE $M \gets M + 1$
        \ENDFOR
    \ENDFOR
    \IF{$m = \varnothing$} \STATE break \ENDIF
    \STATE batch Elo update $\hat{u}(x) \gets \hat{u}(x) + K \cdot (S(x) - E(x) \;\; \forall\;\; x \in \mathcal{X}$
    \STATE $t \gets t + 1$
\ENDWHILE
\RETURN $\hat{u}$, $M$
\end{algorithmic}
\label{algo:swiss_elo}
\end{algorithm}

\begin{algorithm}[h!]
\caption{Random pairs then Swiss Tournament}
\begin{algorithmic}[1]
\STATE \textbf{Input:} Number of items $N$, true values $v$, initial ratings $\mathcal{R}_0$, random rounds $r_{\text{rnd}}$, Swiss rounds $r_{\text{swiss}}$, $K$-factor $K$
\STATE \textbf{Output:} Final ratings $\mathcal{R}$, total number of comparisons $M$

\STATE Initialize ratings $\hat{v} \gets \mathcal{R}_0$
\STATE Let $\mathcal{X} \gets$ set of items with $|\mathcal{X}|=N$
\STATE $M \gets 0$\hfill \COMMENT{total comparisons counter}
\FOR{$t=1$ to $r_{\text{rnd}}$}
    \STATE shuffle $\mathcal{X}$, set $m \gets [\,]$
    \FOR{$i=1$ to $N-1$ step $2$}
        \STATE $x_i, x_j \gets \mathcal{X}(i), \mathcal{X}(i+1)$
        \STATE $(S_i, S_j) \in \{(1,0),(0.5,0.5),(0,1)\}$ w.p. $P(x_i \succ x_j)$ from Eq.~\eqref{eq:BT}
        \STATE $(E_i, E_j) = P(x_i \succ x_j), P(x_j \succ x_i)$ from Eq.~\eqref{eq:Elo}
        \STATE $M \gets M+1$
    \ENDFOR
    \STATE batch Elo update $\hat{u}(x) \gets \hat{u}(x) + K \cdot (S(x) - E(x) \;\; \forall\;\; x \in \mathcal{X}$
\ENDFOR
\FOR{$t=1$ to $r_{\text{swiss}}$}
    \STATE Pair items by Swiss rule: $m \gets \text{pair}(\hat{v})$
    \FORALL{$(x_i,x_j)\in m$}
        \STATE $(S_i, S_j)$ and $(E_i,E_j)$ as before
        \STATE $M \gets M+1$
    \ENDFOR
   \STATE batch Elo update $\hat{v}(x) \gets \hat{v}(x) + K \cdot (S(x) - E(x) \;\; \forall\;\; x \in \mathcal{X}$
\ENDFOR

\RETURN $\hat{v}, M$ 
\end{algorithmic} 
\label{algo:rnd_swiss}
\end{algorithm}

\begin{algorithm}[t]
\caption{Swiss Tournament with Mutual Information Gain pairing}
\begin{algorithmic}[1]
\STATE \textbf{Input:} Number of items $N$, true values $v$, initial ratings $\mathcal{R}_0$, max rounds $r_{\max}$, base $K$-factor $K_0$, min $K$-factor $K_{\min}$
\STATE \textbf{Output:} Final ratings $\mathcal{R}$, total number of comparisons $M$

\STATE Initialize ratings $\hat{v} \gets \mathcal{R}_0$
\STATE $s(x) \gets 0$, $g(x) \gets 0 \;\;\forall x \in \mathcal{X}$
\STATE $M \gets 0$ \COMMENT{total comparisons counter}
\STATE Let $\mathcal{X} \gets$ set of items with $|\mathcal{X}| = N$

\FOR{$t = 1$ to $r_{\max}$}
    \STATE $m \gets [\varnothing]$
    \FORALL{$(x_i, x_j) \in \mathcal{X} \;\;|\;\; x_i \neq x_j$}
        \IF{$(x_i,x_j) \notin m$}
            \STATE $I(x_i,x_j) \gets P(x_i \succ x_j) \cdot (1 - P(x_i \succ x_j))$ Eq.~\eqref{eq:BT}
            \STATE $m \gets m \cup  [I, x_i, x_j]$
        \ENDIF
    \ENDFOR
    \IF{$m = \varnothing$} \STATE break \ENDIF
    \STATE $m \gets \text{sort}(m,I)$
    \FORALL{$(x_i, x_j) \in m$}
        \STATE $(S_i, S_j) \in \{(1,0),(0.5,0.5),(0,1)\}$ w.p. $P(x_i \succ x_j)$ Eq.~\eqref{eq:BT}
        \STATE $(E_i, E_j) = P(x_i \succ x_j), P(x_j \succ x_i)$ from Eq.~\eqref{eq:Elo}
        \STATE $M \gets M+1$
    \ENDFOR
    \STATE batch Elo update $\hat{v}(x) \gets \hat{v}(x) + K \cdot (S(x) - E(x) \;\; \forall\;\; x \in \mathcal{X}$
    \STATE $s(x) \gets \hat{v}(x)$
\ENDFOR
\RETURN $\hat{v}, M$
\end{algorithmic}
\label{algo:swiss_infogain}
\end{algorithm}

%% file: xirl.bib
@article{hilton1986knowledge,
  title={Knowledge-based causal attribution: The abnormal conditions focus model.},
  author={Hilton, Denis J and Slugoski, Ben R},
  journal={Psychological review},
  volume={93},
  number={1},
  pages={75},
  year={1986},
  publisher={American Psychological Association}
}

@article{ibarz2018reward,
  title={Reward learning from human preferences and demonstrations in atari},
  author={Ibarz, Borja and Leike, Jan and Pohlen, Tobias and Irving, Geoffrey and Legg, Shane and Amodei, Dario},
  journal={NeurIPS},
  volume={31},
  year={2018}
}

@article{christiano2017deep,
  title={Deep reinforcement learning from human preferences},
  author={Christiano, Paul F and Leike, Jan and Brown, Tom and Martic, Miljan and Legg, Shane and Amodei, Dario},
  journal={NeurIPS},
  volume={30},
  year={2017}
}

@article{ziegler2019fine,
  author       = {Daniel M. Ziegler and
                  Nisan Stiennon and
                  Jeffrey Wu and
                  Tom B. Brown and
                  Alec Radford and
                  Dario Amodei and
                  Paul F. Christiano and
                  Geoffrey Irving},
  title        = {Fine-Tuning Language Models from Human Preferences},
  journal      = {CoRR},
  volume       = {abs/1909.08593},
  year         = {2019},
  url          = {http://arxiv.org/abs/1909.08593},
  eprinttype    = {arXiv},
  eprint       = {1909.08593},
  bibsource    = {dblp computer science bibliography, https://dblp.org}
}

@article{ouyang2022training,
  title={Training language models to follow instructions with human feedback},
  author={Ouyang, Long and Wu, Jeffrey and Jiang, Xu and Almeida, Diogo and Wainwright, Carroll and Mishkin, Pamela and Zhang, Chong and Agarwal, Sandhini and Slama, Katarina and Ray, Alex and others},
  journal={Advances in neural information processing systems},
  volume={35},
  pages={27730--27744},
  year={2022}
}

@article{casper2023open,
  title={Open Problems and Fundamental Limitations of Reinforcement Learning from Human Feedback},
  author={Casper, Stephen and Davies, Xander and Shi, Claudia and Krendl Gilbert, Thomas and Scheurer, J{\'e}r{\'e}my and Rando Ramirez, Javier and Freedman, Rachel and Korbak, Tomasz and Lindner, David and Freire, Pedro and others},
  journal={Transactions on Machine Learning Research},
  year={2023},
  publisher={OpenReview}
}

@article{Wang2024ACS,
  title={A Comprehensive Survey of LLM Alignment Techniques: RLHF, RLAIF, PPO, DPO and More},
  author={Zhichao Wang and Bin Bi and Shiva K. Pentyala and Kiran Ramnath and Sougata Chaudhuri and Shubham Mehrotra and Zixu Zhu and Xiang-Bo Mao and Sitaram Asur and Na Cheng},
  journal={ArXiv},
  year={2024},
  volume={abs/2407.16216},
  url={https://api.semanticscholar.org/CorpusID:271334000}
}

@article{hua2017swiss,
    author = {Hua, Christopher},
    journal = {University of Pennsylvania},
    year = {2017},
    month = {01},
    title = {The Swiss Tournament Model}
}

@book{mclean1995classics,
  title={Classics of Social Choice},
  author={McLean, I. and Urken, A.B. and Hewitt, F.},
  isbn={9780472104505},
  lccn={94024390},
  url={https://books.google.ie/books?id=0QPv9cg3g-sC},
  year={1995},
  publisher={University of Michigan Press}
}

@article{Bradley1952RankAO,
  title={Rank Analysis of Incomplete Block Designs: I. The Method of Paired Comparisons},
  author={Ralph Allan Bradley and Milton E. Terry},
  journal={Biometrika},
  year={1952},
  volume={39},
  pages={324},
  url={https://api.semanticscholar.org/CorpusID:125209808}
}

@book{elo1978rating,
  title={The Rating of Chessplayers, Past and Present},
  author={Elo, A.E.},
  isbn={9780668047210},
  lccn={78024077},
  url={https://books.google.ie/books?id=8pMnAQAAMAAJ},
  year={1978},
  publisher={Arco Pub.}
}

@article{saari1996copeland,
  title={The copeland method: I.: Relationships and the dictionary},
  author={Saari, Donald G and Merlin, Vincent R},
  journal={Economic theory},
  volume={8},
  number={1},
  pages={51--76},
  year={1996},
  publisher={Springer}
}

@article{copeland1951reasonable,
  title={A ‘reasonable’social welfare function, Seminar on mathematics in social sciences, University of Michigan},
  author={Copeland, A},
  journal={Cited indirectly from its mention by Luce and Raiffa (1957)},
  pages={358},
  year={1951}
}

@inproceedings{dai2024mapping,
  title={Mapping Social Choice Theory to RLHF},
  author={Dai, Jessica and Fleisig, Eve},
  booktitle={ICLR 2024 Workshop on Reliable and Responsible Foundation Models}
}

@inproceedings{maystre2017quicksort,
  title={Just sort it! A simple and effective approach to active preference learning},
  author={Maystre, Lucas and Grossglauser, Matthias},
  booktitle={International Conference on Machine Learning},
  pages={2344--2353},
  year={2017},
  organization={PMLR}
}

@article{shams2011maximum,
  title={Maximum of second-order Taylor approximation of entropy},
  author={Shams, S},
  journal={Applied Mathematical Sciences},
  volume={5},
  number={64},
  pages={3191--3200},
  year={2011}
}
